\documentclass[useAMS,usenatbib]{mn2e}
\usepackage{graphicx}
\usepackage{braket}
\usepackage{hyperref}
\usepackage{color}
\usepackage{multirow}

\title[Analysis of gamma-ray burst duration distribution]{Analysis of gamma-ray burst duration distribution using mixtures of skewed distributions}
\author[M. Tarnopolski]{M. Tarnopolski\thanks{E-mail:
mariusz.tarnopolski@uj.edu.pl}\\
Astronomical Observatory, Jagiellonian University, Orla 171, 30-244 Krak\'{o}w, Poland}
\begin{document}

\date{\today}

\pagerange{\pageref{firstpage}--\pageref{lastpage}} \pubyear{2015}

\maketitle

\label{firstpage}

\begin{abstract}
Two classes of GRBs have been confidently identified thus far and are prescribed to different physical scenarios -- NS-NS or NS-BH mergers, and collapse of massive stars, for short and long GRBs, respectively. A third, intermediate in duration class, was suggested to be present in previous catalogs, such as BATSE and {\it Swift}, based on statistical tests regarding a mixture of two or three log-normal distributions of $T_{90}$. However, this might possibly not be an adequate model. This paper investigates whether the distributions of $\log T_{90}$ from BATSE, {\it Swift}, and {\it Fermi} are described better by a mixture of skewed distributions rather than standard Gaussians. Mixtures of standard normal, skew-normal, sinh-arcsinh and alpha-skew-normal distributions are fitted using a maximum likelihood method. The preferred model is chosen based on the Akaike information criterion. It is found that mixtures of two skew-normal or two sinh-arcsinh distributions are more likely to describe the observed duration distribution of {\it Fermi} than a mixture of three standard Gaussians, and that mixtures of two sinh-arcsinh or two skew-normal distributions are models competing with the conventional three-Gaussian in the case of BATSE and {\it Swift}. Based on statistical reasoning, and it is shown that other phenomenological models may describe the observed {\it Fermi}, BATSE, and {\it Swift} duration distributions at least as well as a mixture of standard normal distributions, and the existence of a third (intermediate) class of GRBs in {\it Fermi} data is rejected.
\end{abstract}


\section{Introduction}\label{intro}

\citet{mazets} first pointed out hints for a bimodal distribution of $T_b$ (taken to be the time interval within which fall $80-90\%$ of the measured GRB's intensity) drawn for 143 events detected in the KONUS experiment. A bimodal structure in the distribution of durations $T_{90}$ (time interval from 5\% to 95\% of the accumulated fluence) in BATSE (Burst Alert and Transient Source Explorer, onboard the Compton Gamma-Ray Observatory, \citealt{meegan92}) 1B dataset, based on which GRBs are nowadays commonly classified into short ($T_{90}<2\,{\rm s}$) and long ($T_{90}>2\,{\rm s}$) classes, was also found \citep{kouve}.While generally short GRBs are of merger origin and long ones come from collapsars, this classification is imperfect due to a large overlap in duration distributions of the two populations \citep{lu,bromberg,bromberg2,shahmoradi,shahmoradi2,tarnopolski2,tarnopolski3}.

\citet{horvath98} discovered a third peak in the duration distribution, located between the short and long ones, in the BATSE 3B catalog, and using multivariate clustering procedures independently the same conclusion was arrived at by \citet{mukh}. The statistical existence of the intermediate class was supported \citep{horvath02} with the use of BATSE 4B data. The evidence for a third normal component in $\log T_{90}$ was found also in {\it Swift}/BAT data \citep{horvath08,zhang,huja,horvath10}. Other datasets, i.e. {\it RHESSI} \citep{ripa,ripa2} and {\it BeppoSAX} \citep{horvath09}, were both in agreement with earlier results regarding the bimodal distribution, and the detection of a third component was established on a lower, compared to BATSE and {\it Swift}, significance level. Hence, four different satellites provided hints about the existence of a third class of GRBs. Contrary to this, durations as observed by {\it INTEGRAL} have a unimodal distribution, which extends to the shortest timescales as a powerlaw \citep{savchenko}. Interestingly, a re-examination of the BATSE current catalog and {\it Swift} dataset \citep{zitouni}, showed that a mixture of three Gaussians fits the {\it Swift} data better than a two-Gaussian, while in the BATSE case statistical tests did not support the presence of a third component\footnote{Adding parameters to a nested  model always results in a better fit (in the sense of a lower $\chi^2$ or a higher maximum log-likelihood, $\mathcal{L}$) due to more freedom given to the model to follow the data, i.e. due to introducing more free parameters. The important question is whether this improvement is statistically significant, and whether the model is justified\label{fn1}.}.

Only one dataset (BATSE 3B) was truly trimodal in the sense of having three peaks (i.e., three local maxima). In the rest (i.e., BATSE 4B and current, {\it Swift}, {\it RHESSI} and {\it BeppoSAX}) a three-Gaussian was found to follow the observations better than a two-Gaussian, but those fits yielded only two peaks, so despite statistical analyses support the presence of a third normal {\it component}, the existence of a third physical {\it class} is not confirmed and may be ascribed to $\log T_{90}$ being described by a distribution different than a mixture of Gaussians, particularly a mixture of skewed distributions \citep{tarnopolski}.

Latest numerous release is due to {\it Fermi}/GBM observations \citep{gruber,kienlin} and consists of $\sim 1600$ GRBs with computed durations $T_{90}$. Up to date, to the best of the author's knowledge, except \citet{tarnopolski}, only \citet{horvath12}, \citet{zhang2} and \citet{qin} conducted research on a {\it Fermi} subsample, consisting of 425 bursts, from the first release of the catalog.

It was proposed \citep{tarnopolski}, in the light of \citet{zitouni}, who suggested that the non-symmetry of the $\log T_{90}$ distributions is due to a non-symmetric distribution of the envelope masses of the progenitors, that a mixture of skewed distributions might be phenomenologically a better model than the commonly applied mixture of standard Gaussians. The aim of this paper is to examine whether mixtures of various skewed distributions (skew-normal, sinh-arcsinh and alpha-skew-normal) describe the duration distribution better than a mixture of standard Gaussians. Particularly, it is verified whether two-component mixtures of skewed distributions might challenge a commonly applied three-Gaussian model. If this is shown to be true, the existence of the intermediate class in the duration distribution will be questioned.

Because the $T_{90}$ distribution is detector dependent \citep{nakar,tarnopolski2}, the analysis herein is not restricted to the {\it Fermi} dataset as it was in \citep{tarnopolski}, but also the BATSE and {\it Swift} data are examined. These three datasets have been fitted to date with a mixture of standard Gaussians, but to the best of the author's knowledge no other types of distributions were applied to the observed $T_{90}$ distributions. It may happen that due to instrument specification a three-component distribution might be a better description for some datasets, while for others a two-component one will be sufficient \citep{zitouni}.

This article is organized as follows. Section~\ref{data} describes the datasets, fitting method, the properties of the distributions examined, and the method of assessing the goodness of fit. In Section~\ref{res} the results are presented. Section~\ref{disc} is devoted to discussion, and Section~\ref{conc} gives concluding remarks. The computer algebra system \textsc{Mathematica}\textsuperscript{\textregistered} v10.0.2 is applied throughout this paper.

\section{Datasets, methods and distributions}\label{data}

\subsection{Samples}\label{data1}

The datasets\footnote{All accessed on April 29, 2015.} from {\it Fermi}\footnote{\url{http://heasarc.gsfc.nasa.gov/W3Browse/fermi/fermigbrst.html} \citep{gruber,kienlin}}, BATSE\footnote{\url{http://gammaray.msfc.nasa.gov/batse/grb/catalog/current} \citep{meegan98,paciesas}}, and {\it Swift}\footnote{\url{http://swift.gsfc.nasa.gov/archive/grb\_table/} \citep{gehrels}} are considered herein. The BATSE current catalog consists of 2041 GRBs, and the {\it Swift} dataset contains 914 events. {\it Fermi} observed 1596 GRBs, but a dataset of 1593 GRBs is used. Three durations that stand out (two shortest and one longest) were treated as outliers and excluded due to their significant separation from the remaining durations and a possibility of a strong influence on the outcome, especially on the tails of the fitted distributions (if the data are binned according to any well established rule (see Section~\ref{data2}), the bins containing these three values are separated by empty bins from the rest of the distribution). Whereas the durations $T_{90}$ are approximately log-normally distributed, herein their decimal logarithms, $\log T_{90}$'s, are employed; for simplicity they will be referred to as \textit{durations} too, and whenever a phrase \textit{normal distribution of durations} is used, it is understood in the sense of \textit{normal distribution of logarithms of durations ($\log T_{90}$)} or, equivalently, \textit{log-normal distribution of durations $T_{90}$}. This notion applies also to other distributions examined throughout this paper.

The RHESSI and BeppoSAX datasets are not examined here for the following reasons: {\it i}) RHESSI has no GRB trigerring and only consists of GRBs observed by other satellites; {\it ii}) RHESSI is a relatively small dataset (427 GRBs, \citet{ripa,ripa2}); {\it iii}) BeppoSAX, due to its relatively long ($1\,{\rm s}$) short integration time \citep{horvath09}, does not contain many short GRBs.

\subsection{Fitting method}\label{data2}

Two standard fitting techniques are commonly applied: $\chi^2$ fitting and maximum likelihood (ML) method. For the first, data need to be binned, and despite various binning rules are known (e.g. Freedman-Diaconis, Scott, Knuth etc.), they still leave place for ambiguity, as it might happen that the fit may be statistically significant on a given significance level for a number of binnings \citep{huja2,koen,tarnopolski}. The ML method is not affected by this issue and is therefore applied herein. However, for display purposes, the binnings were chosen based on the Knuth rule.

Having a distribution with a probability density function (PDF) given by $f=f(x;\theta)$ (possibly a mixture), where $\theta=\left\{\theta_i\right\}_{i=1}^p$ is a set of $p$ parameters, the log-likelihood function is defined as
\begin{equation}
\mathcal{L}_p(\theta)=\sum\limits_{i=1}^N\ln f(x_i;\theta),
\label{eq1}
\end{equation}
where $\left\{x_i\right\}_{i=1}^N$ are the datapoints from the sample to which a distribution is fitted. The fitting is performed by searching a set of parameters $\hat{\theta}$ for which the log-likelihood is maximized \citep{kendall}. When nested models are considered, the maximal value of the log-likelihood function $\mathcal{L}_{\rm max}\equiv\mathcal{L}_p(\hat{\theta})$ increases when the number of parameters $p$ increases.

\subsection{Distributions and their properties}\label{data3}

The following distributions are considered.

A mixture of $k$ standard normal (Gaussian) $\mathcal{N}(\mu,\sigma^2)$ distributions:
\begin{equation}
\begin{array}{l}
f^{(\mathcal{N})}_k(x) = \sum\limits_{i=1}^k A_i \varphi\left(\frac{x-\mu_i}{\sigma_i}\right) \\
\textcolor{white}{f^{(\mathcal{N})}_k(x)} = \sum\limits_{i=1}^k \frac{A_i}{\sqrt{2\pi}\sigma_i}\exp\left(-\frac{(x-\mu_i)^2}{2\sigma_i^2}\right),
\end{array}
\label{eq5}
\end{equation}
being described by $3k-1$ free parameters: $k$ pairs $(\mu_i,\sigma_i)$ and $k-1$ weights $A_i$, satysfying $\sum_{i=1}^k A_i=1$. Skewness of each component is $\gamma_1^{(\mathcal{N})}=0$.

A mixture of $k$ skew normal (SN) distributions \citep{ohagan,azzalini}:
\begin{equation}
\begin{array}{l}
f^{(\mathcal{SN})}_k(x) = \sum\limits_{i=1}^k 2A_i\varphi\left(\frac{x-\mu_i}{\sigma_i}\right)\Phi\left(\alpha_i\frac{x-\mu_i}{\sigma_i}\right) \\
\textcolor{white}{f^{(\mathcal{SN})}_k(x)} = \sum\limits_{i=1}^k \frac{2A_i}{\sqrt{2\pi}\sigma_i}\exp\left(-\frac{(x-\mu_i)^2}{2\sigma_i^2}\right)\times \\
\textcolor{white}{f^{(\mathcal{SN})}_k(x)  = \sum\limits_{i=1}^k} \times\frac{1}{2}\left[1+\textrm{erf}\left(\alpha_i\frac{x-\mu_i}{\sqrt{2}\sigma_i}\right)\right],
\end{array}
\label{eq6}
\end{equation}
described by $4k-1$ parameters. Skewness of an SN distribution is
\[
\gamma_1^{(\mathcal{SN})}=\frac{4-\pi}{2}\frac{\left(\zeta\sqrt{2/\pi}\right)^3}{\left(1-2\zeta^2/\pi\right)^{3/2}},
\]
where $\zeta=\frac{\alpha}{\sqrt{1+\alpha^2}}$, hence the skewness $\gamma_1^{(\mathcal{SN})}$ is solely based on the shape parameter $\alpha$, and is limited roughly to the interval $\left(-1,1\right)$. The mean is given by $\mu+\sigma\zeta\sqrt{\frac{2}{\pi}}$. When $\alpha=0$, the SN distribution is reduced to a standard Gaussian $\mathcal{N}(\mu,\sigma^2)$ due to $\Phi(0)=1/2$.

A mixture of $k$ sinh-arcsinh (SAS) distributions \citep{jones}:
\begin{equation}
\begin{array}{l}
f^{(\mathcal{SAS})}_k(x) = \sum\limits_{i=1}^k \frac{A_i}{\sigma_i}\left[1+\left(\frac{x-\mu_i}{\sigma_i}\right)^2\right]^{-\frac{1}{2}} \times\\
\textcolor{white}{f^{(\mathcal{SAS})}_k(x) = \sum\limits_{i=1}^k} \times\beta_i \cosh\left[\beta_i\sinh^{-1}\left(\frac{x-\mu_i}{\sigma_i}\right)-\delta_i\right]\times \\
\textcolor{white}{f^{(\mathcal{SAS})}_k(x) = \sum\limits_{i=1}^k} \times\exp\left[-\frac{1}{2}\sinh\left[\beta_i\sinh^{-1}\left(\frac{x-\mu_i}{\sigma_i}\right)-\delta_i\right]^2\right],
\end{array}
\label{eq7}
\end{equation}
being described by $5k-1$ parameters. It turns out that skewness of the SAS distribution increases with increasing $\delta$, positive skewness corresponding to $\delta>0$. Tailweight decreases with increasing $\beta$, $\beta<1$ yielding heavier tails than the normal distribution, and $\beta>1$ yielding lighter tails. With $\delta=0$ and $\beta=1$, the SAS distribution reduces to a standard Gaussian, $\mathcal{N}(\mu,\sigma^2)$. Skewness of a SAS distribution is
\[
\gamma_1^{(\mathcal{SAS})}=\frac{1}{4}\left[ \sinh\left(\frac{3\delta}{\beta}\right)P_{3/\beta} - 3\sinh\left(\frac{\delta}{\beta}\right)P_{1/\beta} \right],
\]
where
\[
P_q=\frac{e^{1/4}}{\sqrt{8\pi}}\left[ K_{(q+1)/2}(1/4) + K_{(q-1)/2}(1/4) \right].
\]
Here, $K$ is the modified Bessel function of the second kind. The mean is given by $\mu+\sigma\sinh(\delta/\beta)P_{1/\beta}$.

A mixture of $k$ alpha-skew-normal (ASN) distributions \citep{elal}:
\begin{equation}
\begin{array}{l}
f^{(\mathcal{ASN})}_k(x) = \sum\limits_{i=1}^k A_i\frac{\left(1-\alpha_i\frac{x-\mu_i}{\sigma_i}\right)^2+1}{2+\alpha_i^2} \varphi\left(\frac{x-\mu_i}{\sigma_i}\right) \\
\textcolor{white}{f^{(\mathcal{ASN})}_k(x)} = \sum\limits_{i=1}^k A_i\frac{\left(1-\alpha_i\frac{x-\mu_i}{\sigma_i}\right)^2+1}{2+\alpha_i^2}\frac{1}{\sqrt{2\pi}\sigma_i}\exp\left(-\frac{(x-\mu_i)^2}{2\sigma_i^2}\right),
\end{array}
\label{eq8}
\end{equation}
described by $4k-1$ parameters. Skewness of an ASN distribution is
\[
\gamma_1^{(\mathcal{ASN})}=\frac{12\alpha^5+8\alpha^3}{(3\alpha^4+4\alpha^2+4)^{3/2}},
\]
and is limited roughly to the interval $(-0.811,0.811)$. The mean is given by $\mu-\frac{2\alpha\sigma}{2+\alpha^2}$. For $\alpha\in(-1.34,1.34)$ the distribution is unimodal, and bimodal otherwise.

\subsection{Assessing the likelihood of the fits}\label{data4}

If one has two fits such that $\mathcal{L}_{p_2,{\rm max}} > \mathcal{L}_{p_1,{\rm max}}$, then twice their difference, $2\Delta\mathcal{L}_{\rm max}=2(\mathcal{L}_{p_2,{\rm max}}-\mathcal{L}_{p_1,{\rm max}})$, is distributed like $\chi^2(\Delta p)$, where $\Delta p=p_2-p_1>0$ is the difference in the number of parameters \citep{kendall,horvath02}. If a $p$-value associated with the value of $\chi^2(\Delta p)$ does not exceed the significance level $\alpha$, one of the fits (with higher $\mathcal{L}_{\rm max}$) is statistically better than the other. For instance, for a 2-G and a 3-G, $\Delta p=3$, and despite that, according to Footnote~\ref{fn1}, $\mathcal{L}_{\rm max,\,3-G} > \mathcal{L}_{\rm max,\,2-G}$ holds always, twice their difference provides a decisive $p$-value.

It is crucial to note that it follows from the above description that this method is not suitable for situations when the model with the higher $\mathcal{L}_{\rm max}$ has fewer parameters, i.e. $\Delta\mathcal{L}_{\rm max}>0$ and $\Delta p<0$. Moreover, while all of the skewed distributions considered herein contain the standard Gaussian as their special case, what makes them nested models, but e.g. the SAS and SN distributions are not nested, hence no direct comparison can be performed for them with this approach.

For nested as well as non-nested models, the Akaike information criterion ($AIC$) \citep{akaike,burnham,biesiada,liddle} may be applied. The $AIC$ is defined as
\begin{equation}
AIC=2p-2\mathcal{L}_{\rm max}.
\label{eq3}
\end{equation}
A preferred model is the one that minimizes $AIC$. The formulation of $AIC$ penalizes the use of an excessive number of parameters, hence discourages overfitting. It prefers models with fewer parameters, as long as the others do not provide a substantially better fit. The expression for $AIC$ consists of two competing terms: the first measuring the model complexity (number of free parameters) and the second measuring the goodness of fit (or more precisely, the lack of thereof). Among candidate models with $AIC_i$, let $AIC_{\rm min}$ denote the smallest. Then,
\begin{equation}
Pr_i=\exp\left(-\frac{\Delta_i}{2}\right),
\label{eq4}
\end{equation}
where $\Delta_i=AIC_i-AIC_{\rm min}$, can be interpreted as the relative (compared to $AIC_{\rm min}$) probability that the $i$-th model minimizes the $AIC$.

What is essential in assesing the goodness of a fit in the $AIC$ method is the difference, $\Delta_i=AIC_i-AIC_{\rm min}$, not the absolute value\footnote{The $AIC$ value contains scaling constants coming from the log-likelihood $\mathcal{L}$, and so $\Delta_i$ are free of such constants \citep{burnham}. One might consider $\Delta_i=AIC_i-AIC_{\rm min}$ a rescaling transformation that forces the best model to have $\Delta_{\rm min}:=0$.} of an $AIC_i$. If $\Delta_i<2$, then there is substantial support for the $i$-th model (or the evidence against it is worth only a bare mention), and the proposition that it is a proper description is highly probable. If $2<\Delta_i<4$, then there is strong support for the $i$-th model. When $4<\Delta_i<7$, there is considerably less support, and models with $\Delta_i>10$ have essentially no support \citep{burnham,biesiada}. It is important to note that when two models with similar $\mathcal{L}_{\rm max}$ are considered, the $\Delta_i$ depends solely on the number of parameters due to the $2p$ term in Eq.~(\ref{eq3}). Hence, when $\Delta_i/2\Delta p<1$, the relative improvement is due to actual improvement of the fit, not to increasing the number of parameters only.

Finally, $AIC$ tries to select a model that most adequately describes reality (in the form of the data under examination). This means that in fact the model being a real description of the data is never considered.

\section{Results}\label{res}

\subsection{Finding the number of components -- standard Gaussian case}

\begin{figure}
\begin{center}
\includegraphics[width=0.99\linewidth]{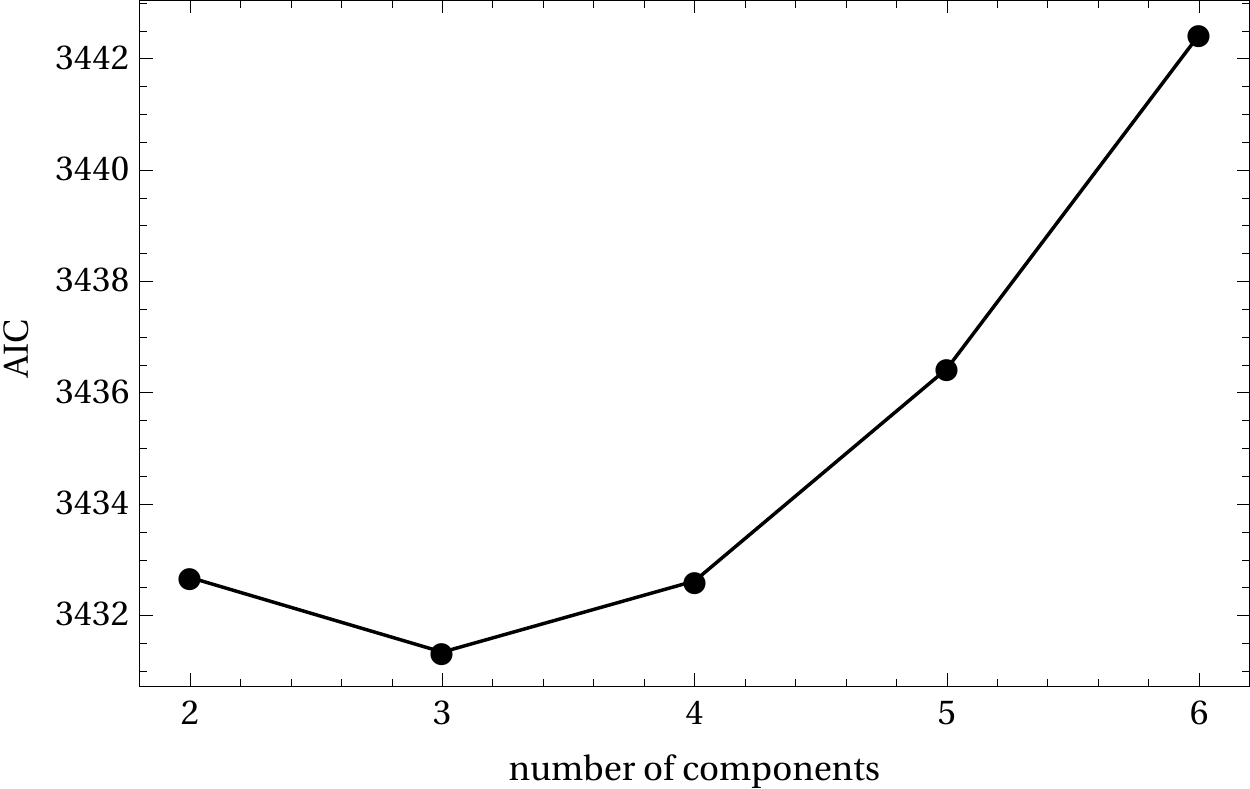}
\end{center}
\caption{$AIC$ vs. number of components in a mixture of standard normal distributions. The minimal value corresponds to a three-Gaussian.}
\label{fig_AICplot}
\end{figure}

First, a mixture of standard Gaussians given by Eq.~(\ref{eq5}) is fitted using the ML method, i.e. maximizing $\mathcal{L}$ given by Eq.~(\ref{eq1}). The mixtures range from $k=2$ to $k=6$ components. The $AIC$ is calculated by means of Eq.~(\ref{eq3}). The preferred model is the one with the lowest $AIC$, and it follows from Figure~\ref{fig_AICplot} that among the Gaussian models examined, a mixture of three components is the most plausible to describe the observed distribution of {\it Fermi} $\log T_{90}$. The same conclusion is drawn for the BATSE and {\it Swift} datasets. Hence, as it is expected that the other PDFs [SN, SAS and ASN given by Eq.~(\ref{eq6})--(\ref{eq8})] will be more flexible in fitting the data, the forthcoming analysis is restricted to two or three components for distributions being a mixture of unimodal PDFs (SN and SAS), and to one, two, or three components for ASN, as its one bimodal component may turn out to follow the data well enough.

\subsection{Fitting the distributions}

\subsubsection{{\it Fermi}}

\begin{figure}
\begin{center}
\includegraphics[width=0.99\linewidth]{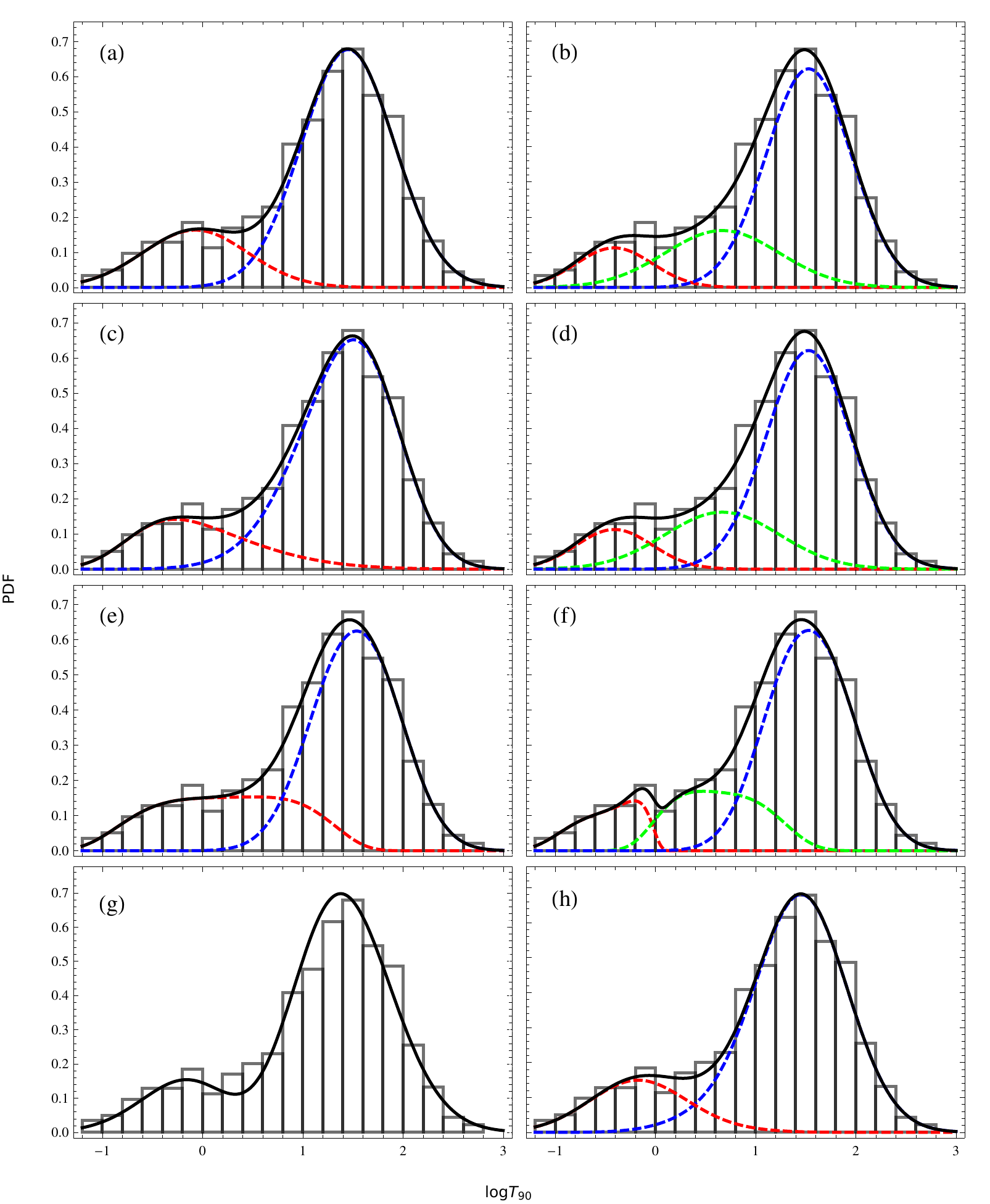}
\end{center}
\caption{Distributions fitted to $\log T_{90}$ data gathered by {\it Fermi}. Color dashed curves are the components of the (black solid) mixture distribution. The panels show a mixture of (a) two standard Gaussians, (b) three standard Gaussians, (c) two skew-normal, (d) three skew-normal, (e) two sinh-arcsinh, (f) three sinh-arcsinh, (g) one alpha-skew-normal, and (h) two alpha-skew-normal distributions.}
\label{fig1}
\end{figure}

\begin{table*}
\caption{Parameters of the fits to the {\it Fermi} data. Label corresponds to labels from Figure~\ref{fig1}. The smallest $AIC$ is marked in bold, and $p$ is the number of parameters in a model.}
\label{tbl1}
\centering
\begin{tabular}{|c c c c c c c c c c c c c c|}
\hline
  Label & Dist. & $i$ & $\mu_i$ & $\sigma_i$ & $\alpha_i$ & $\delta_i$ & $\beta_i$ & $A_i$ & $\mathcal{L}_{\rm max}$ & $AIC$ & $\Delta AIC$ & $Pr$ & $p$ \\
  \hline
\multirow{2}{*}{(a)} & \multirow{2}{*}{2-G} & 1 & $-0.073$ & 0.525 & --- & --- & --- & 0.215 & \multirow{2}{*}{$-1711.342$} & \multirow{2}{*}{3432.683} & \multirow{2}{*}{4.459} & \multirow{2}{*}{0.108} & \multirow{2}{*}{5} \\
    &     & 2 &  1.451 & 0.463 & --- & --- & --- & 0.785 & & & & & \\
  \hline
    &     & 1 & $-0.409$ & 0.379 & --- & --- & --- & 0.107 & & & & & \\
(b) & 3-G & 2 &  0.668 & 0.570 & --- & --- & --- & 0.231 & $-1707.672$ & 3431.343 & 3.119 & 0.210 & 8 \\
    &     & 3 &  1.530 & 0.426 & --- & --- & --- & 0.662 & & & & & \\
  \hline
\multirow{2}{*}{(c)} & \multirow{2}{*}{2-SN} & 1 & $-0.735$ & 0.954 &  2.819 & --- & --- & 0.208 & \multirow{2}{*}{$-1707.112$} & \multirow{2}{*}{\textbf{3428.224}} & \multirow{2}{*}{0} & \multirow{2}{*}{1} & \multirow{2}{*}{7} \\
    &      & 2 &  1.865 & 0.664 & $-1.507$ & --- & --- & 0.792 & & & & & \\
  \hline
    &      & 1 & $-0.415$ & 0.379 &  0.019 & --- & --- & 0.107 & & & & & \\
(d) & 3-SN & 2 &  0.726 & 0.573 & $-0.127$ & --- & --- & 0.231 & $-1707.672$ & 3437.343 & 9.119 & 0.010 & 11 \\
    &      & 3 &  1.515 & 0.426 &  0.044 & --- & --- & 0.662 & & & & & \\
  \hline
\multirow{2}{*}{(e)} & \multirow{2}{*}{2-SAS} & 1 & 1.537 & 0.468 & --- & $-0.014$ & 1.068 & 0.685 & \multirow{2}{*}{$-1706.089$} & \multirow{2}{*}{3430.177} & \multirow{2}{*}{1.953} & \multirow{2}{*}{0.377} & \multirow{2}{*}{9} \\
    &       & 2 & 2.158 & 6.146 & --- & $-2.367$ & 7.756 & 0.315 & & & & & \\
  \hline
    &       & 1 & 0.434 & 1.063 & --- &  0.370 & 2.111 & 0.214 & & & & & \\
(f) & 3-SAS & 2 & 0.473 & 0.402 & --- & $-4.161$ & 2.680 & 0.111 & $-1704.248$ & 3436.497 & 8.273 & 0.016 & 14 \\
    &       & 3 & 1.529 & 0.468 & --- &  0.020 & 1.087 & 0.675 & & & & & \\
  \hline
(g) & 1-ASN & 1 & 0.744 & 0.590 & $-1.712$ & --- & --- & 1 & $-1725.038$ & 3456.077 & 27.853 & $<10^{-6}$ & 3 \\
  \hline
\multirow{2}{*}{(h)} & \multirow{2}{*}{2-ASN} & 1 & 0.087 & 0.499 &  0.535 & --- & --- & 0.186 & \multirow{2}{*}{$-1710.427$} & \multirow{2}{*}{3434.853} & \multirow{2}{*}{6.629} & \multirow{2}{*}{0.036} & \multirow{2}{*}{7} \\
    &       & 2 & 1.150 & 0.483 & $-0.667$ & --- & --- & 0.814 & & & & & \\
  \hline
\end{tabular}
\end{table*}

\begin{figure}
\begin{center}
\includegraphics[width=0.99\linewidth]{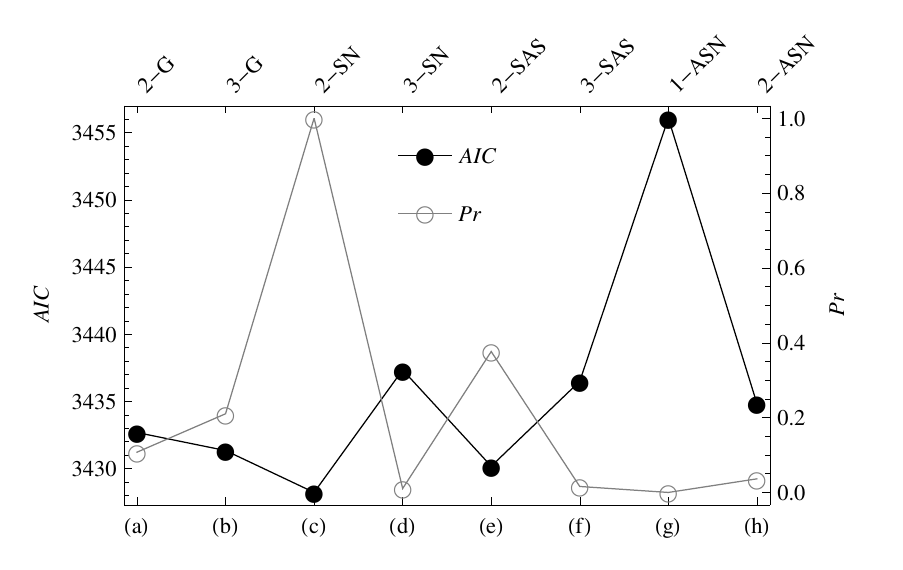}
\end{center}
\caption{$AIC$ and relative probability ($Pr$) for the {\it Fermi} models.}
\label{fig2}
\end{figure}

The following distributions are examined: a two- and three-Gaussian (2-G and 3-G), a two- and three-SN (2-SN and 3-SN), a two- and three-SAS (2-SAS and 3-SAS), a one- and two-ASN (1-ASN and 2-ASN). The results in graphical form are displayed in Figure~\ref{fig1}, whereas the fitted parameters are gathered in Table~\ref{tbl1}, which contains also the values of $\mathcal{L}_{\rm max}$, $AIC$ and relative probability, given by Eq.~(\ref{eq1}), (\ref{eq3}) and (\ref{eq4}), respectively. For completeness, a mixture of three ASN distributions was also fitted to the data, and turnt out to be the worst among the fits obtained, with $AIC=3496.548$ (i.e., higher by about 40 than the highest $AIC$, corresponding to a 1-ASN, from Table~\ref{tbl1}). To visualize the relative goodness-of-fits, the values of $AIC$ and the relative probabilities are shown in Figure~\ref{fig2}. The minimal $AIC$ is obtained by a \mbox{2-SN} distribution. There is also a 37.7\% probability that a 2-SAS distribution describes the data. Both distributions consist of two components and are bimodal. The third lowest $AIC$ was attained by a three-Gaussian distribution with a probability of being correct equal to 21\% (corresponding to $\Delta_{\rm 3-G}=3.119$, which is a somewhat weaker support than the 2-SAS has relative to a 2-SN). While the two-Gaussian exhibits a significant 10.8\% probability of being the correct distribution, it is only the fourth among the eight tested, with considerably less support (i.e., $\Delta_{\rm 2-G}=4.459$). The remaining four (2-ASN, 3-SAS, 3-SN and 1-ASN) have only a few percent of chance for describing the duration distribution, therefore are unlikely to be a proper model.

\subsubsection{BATSE and {\it Swift}}

\begin{figure}
\begin{center}
\includegraphics[width=0.99\linewidth]{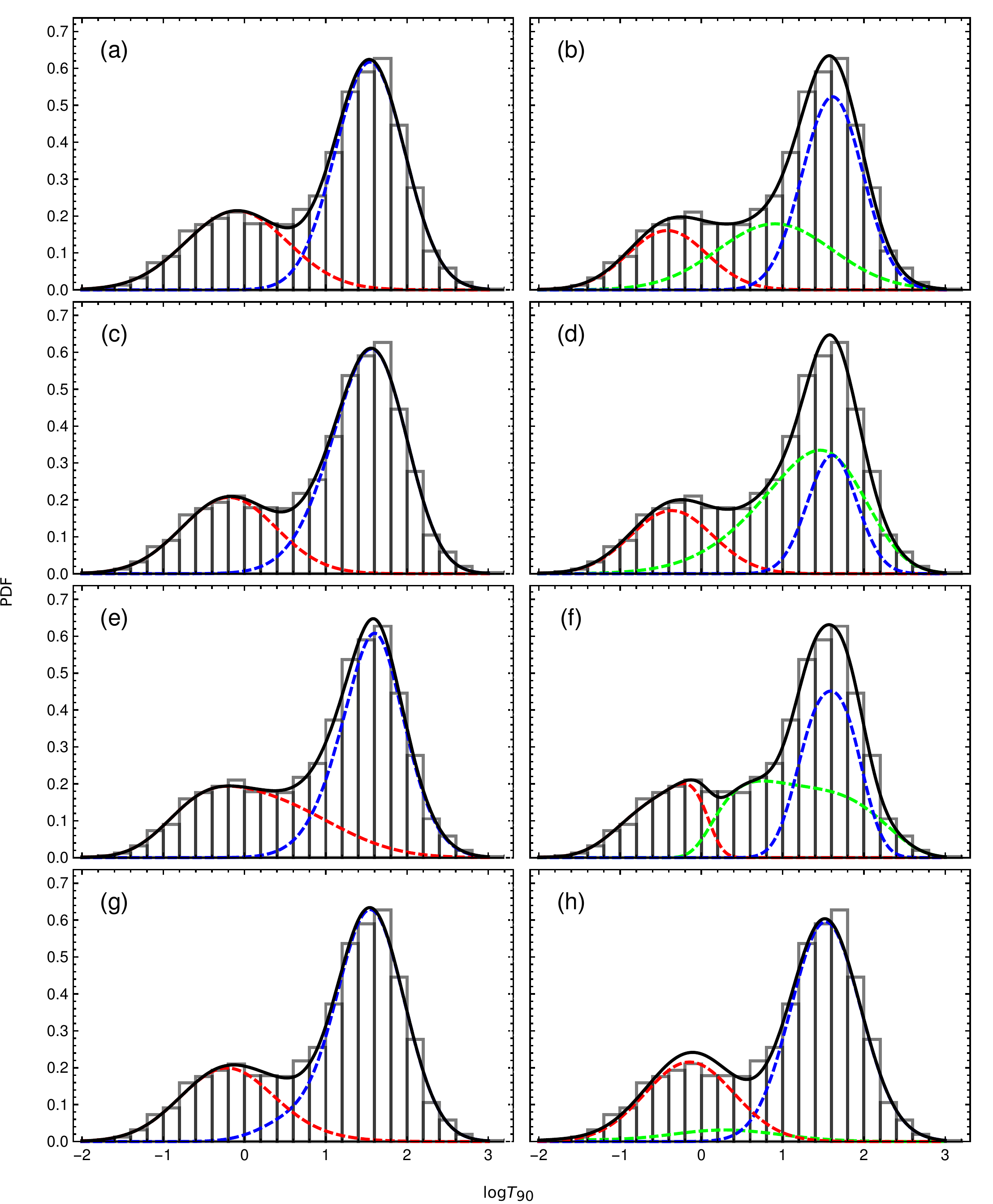}
\end{center}
\caption{Distributions fitted to $\log T_{90}$ data from the BATSE current catalog. Color dashed curves are the components of the (black solid) mixture distribution. The panels show a mixture of (a) two standard Gaussians, (b) three standard Gaussians, (c) two skew-normal, (d) three skew-normal, (e) two sinh-arcsinh, (f) three sinh-arcsinh, (g) two alpha-skew-normal, and (h) three alpha-skew-normal distributions.}
\label{fig3}
\end{figure}

\begin{figure}
\begin{center}
\includegraphics[width=0.99\linewidth]{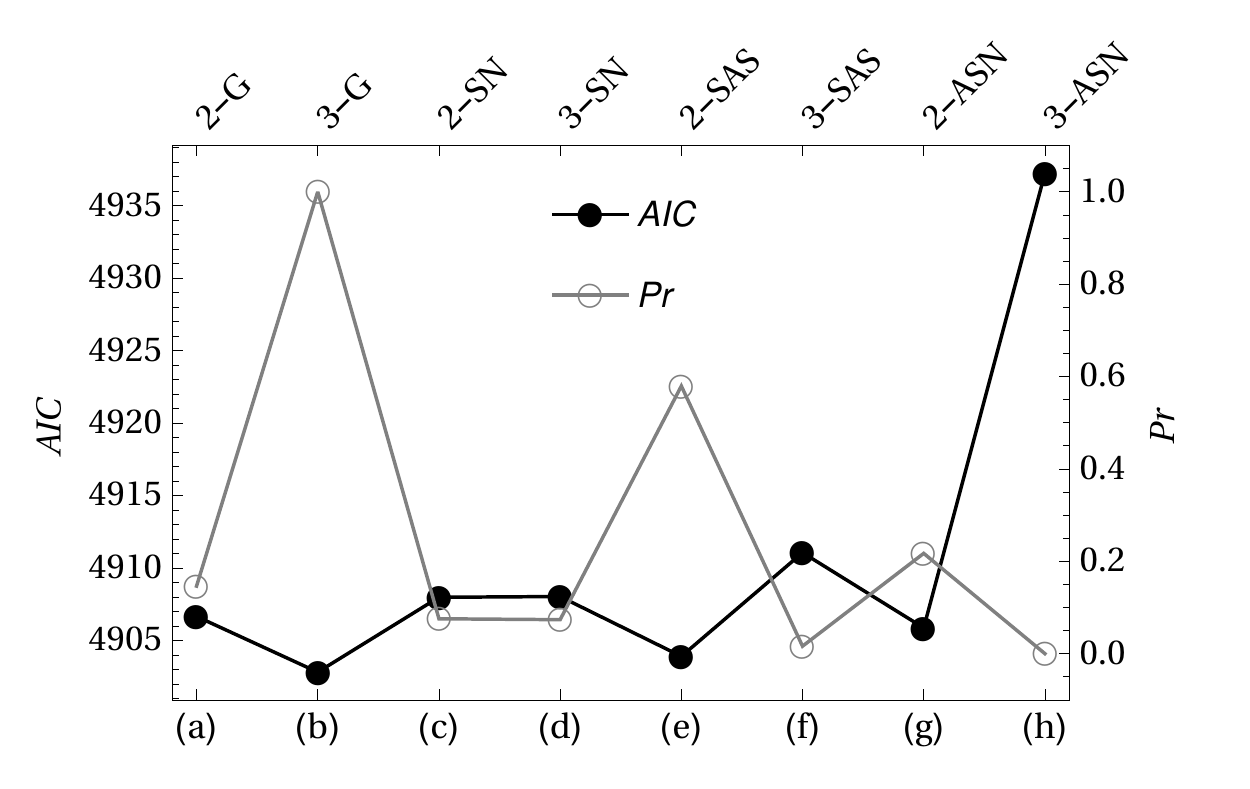}
\end{center}
\caption{$AIC$ and relative probability ($Pr$) for the BATSE models.}
\label{fig4}
\end{figure}

\begin{figure}
\begin{center}
\includegraphics[width=0.99\linewidth]{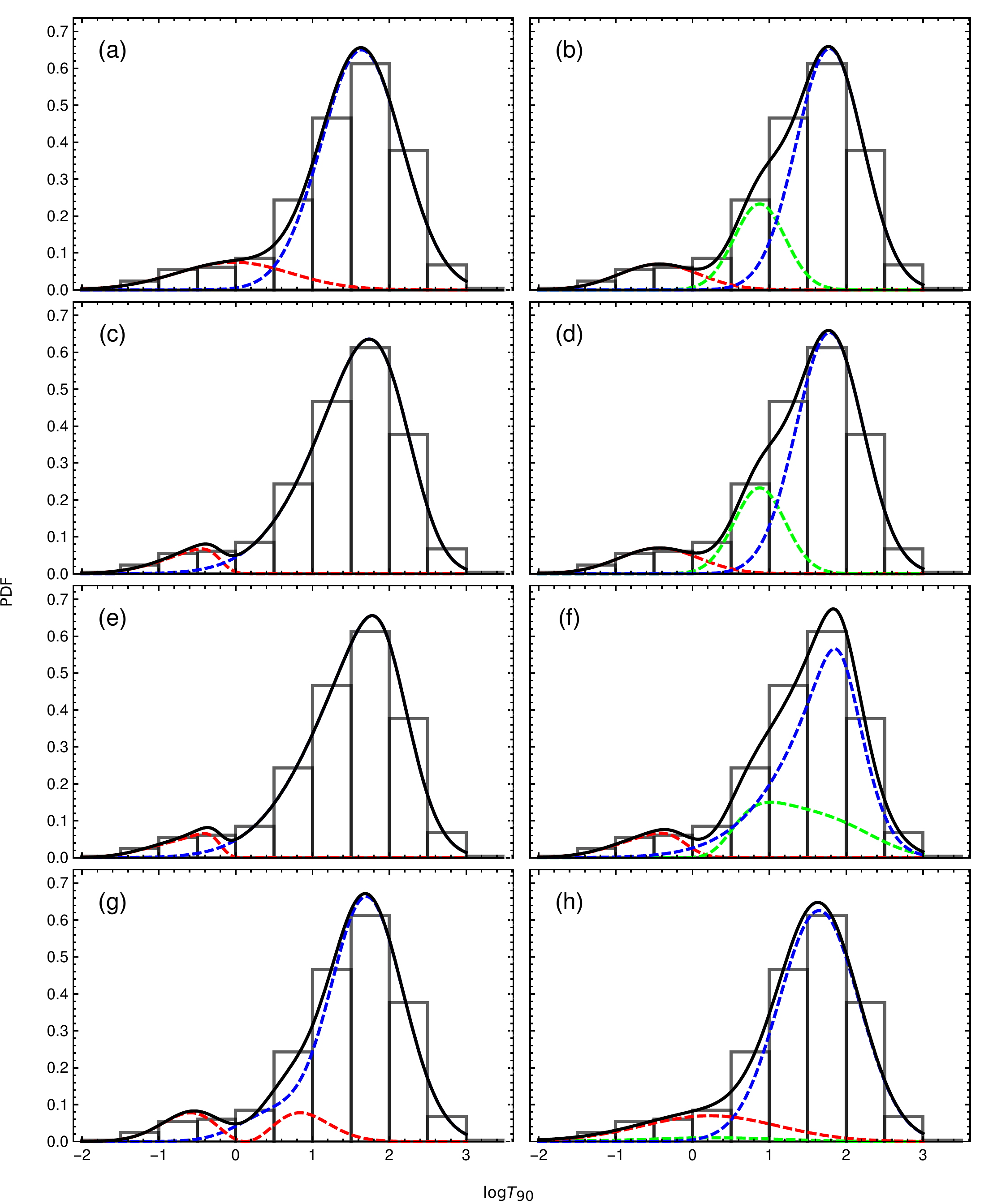}
\end{center}
\caption{Distributions fitted to $\log T_{90}$ data observed by {\it Swift}. Color dashed curves are the components of the (black solid) mixture distribution. The panels show a mixture of (a) two standard Gaussians, (b) three standard Gaussians, (c) two skew-normal, (d) three skew-normal, (e) two sinh-arcsinh, (f) three sinh-arcsinh, (g) two alpha-skew-normal, and (h) three alpha-skew-normal distributions.}
\label{fig5}
\end{figure}

\begin{figure}
\begin{center}
\includegraphics[width=0.99\linewidth]{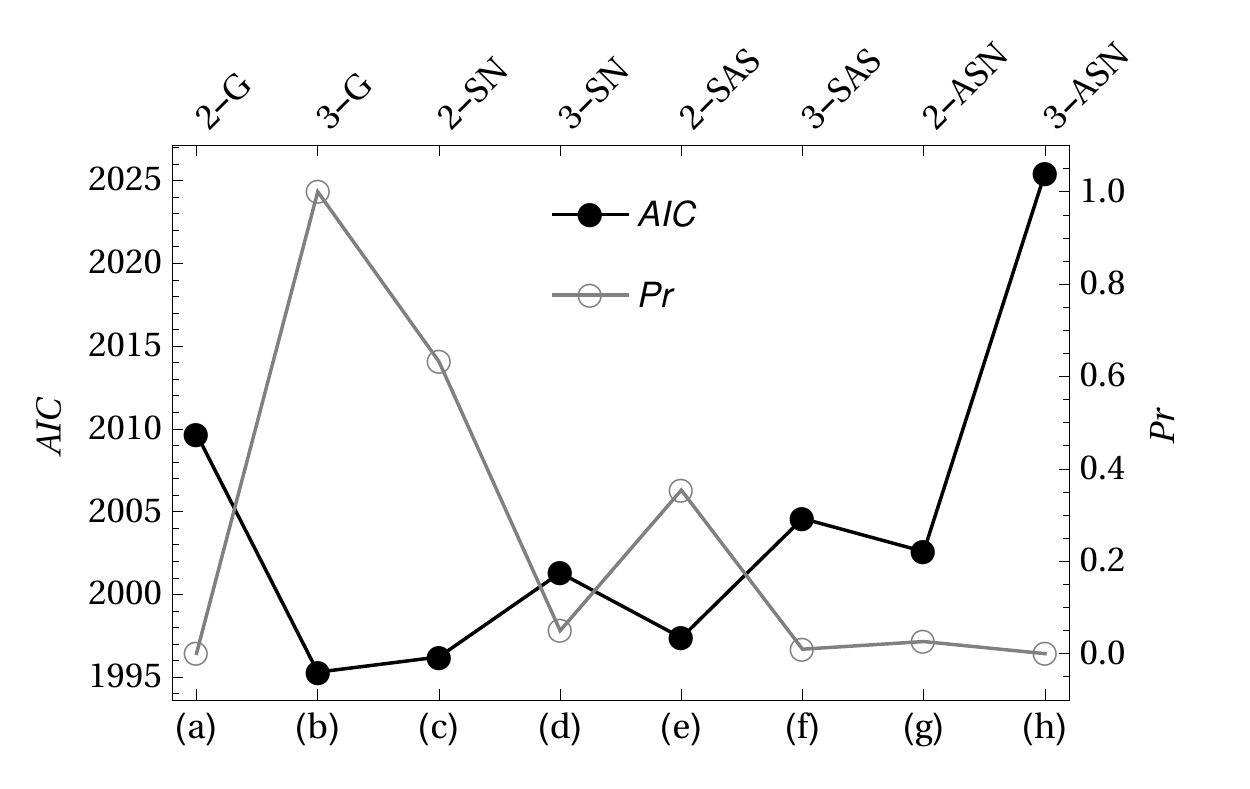}
\end{center}
\caption{$AIC$ and relative probability ($Pr$) for the {\it Swift} models.}
\label{fig6}
\end{figure}

The results are slightly different for the BATSE and {\it Swift} datasets, and are displayed in graphical form in Figures~\ref{fig3}~and~\ref{fig5}. Here, instead of fitting a 1-ASN and a 2-ASN, a 2-ASN and \mbox{a 3-ASN} distributions are fitted, because the 1-ASN yielded an $AIC$ so large that a comparison with other models would be uninsightful\footnote{For BATSE, $AIC_{\rm 1-ASN}=5029.805$, being higher by almost 100 than the highest $AIC$, corresponding to a 3-ASN, and for {\it Swift} \mbox{$AIC_{\rm 1-ASN}=2029.240$}, being by about 4 bigger than the highest $AIC$ (also corresponding to a 3-ASN), and by almost 35 higher than the lowest $AIC$ (attained for a 3-G); compare with Table~\ref{tbl2} and \ref{tbl3}.}. For both samples, the minimal $AIC$ is obtained for a mixture of three standard Gaussians, hence the results of all the previous analyses are confirmed \citep{horvath02,horvath08,zhang,horvath09,huja,huja2,zitouni}. However, for the second best models (2-SAS and 2-SN for BATSE and {\it Swift}, respectively), the $\Delta AIC$ is $\approx 1$, corresponding to a relative probability of 57.9\% and 63.2\% for BATSE and {\it Swift}, respectively (see Table~\ref{tbl2}~and~\ref{tbl3}). This is a substantial support for these two-component models \citep{burnham,biesiada}, hence they cannot be ruled out (see also Figures~\ref{fig4}~and~\ref{fig6}). The next lowest, i.e. third and fourth, $AIC$ for the BATSE data correspond to a 2-ASN and a 2-G, while the {\it Swift} dataset is well described by a 2-SAS or 2-ASN distribution. The rest of the models examined have a relative probability of being a better description of the data than a 3-G distribution less than 10\%. The 3-ASN has a negligible relative probability for both datasets.

\begin{table*}
\caption{Parameters of the fits to the BATSE data. Label corresponds to labels from Figure~\ref{fig3}. The smallest $AIC$ is marked in bold, and $p$ is the number of parameters in a model.}
\label{tbl2}
\centering
\begin{tabular}{|c c c c c c c c c c c c c c|}
\hline
  Label & Dist. & $i$ & $\mu_i$ & $\sigma_i$ & $\alpha_i$ & $\delta_i$ & $\beta_i$ & $A_i$ & $\mathcal{L}_{\rm max}$ & $AIC$ & $\Delta AIC$ & $Pr$ & $p$ \\
  \hline
\multirow{2}{*}{(a)} & \multirow{2}{*}{2-G} & 1 & $-0.095$ & 0.627 & --- & --- & --- & 0.336 & \multirow{2}{*}{$-2448.329$} & \multirow{2}{*}{4906.659} & \multirow{2}{*}{3.844} & \multirow{2}{*}{0.146} & \multirow{2}{*}{5} \\
    &     & 2 &  1.544 & 0.429 & --- & --- & --- & 0.664 & & & & & \\
  \hline
    &     & 1 & $-0.420$ & 0.487 & --- & --- & --- & 0.196 & & & & & \\
(b) & 3-G & 2 &   0.907  & 0.705 & --- & --- & --- & 0.316 & $-2443.407$ & \textbf{4902.815} & 0 & 1 & 8 \\
    &     & 3 &   1.615  & 0.372 & --- & --- & --- & 0.488 & & & & & \\
  \hline
\multirow{2}{*}{(c)} & \multirow{2}{*}{2-SN} & 1 & $-0.193$ & 0.578 &  0.001 & --- & --- & 0.300 & \multirow{2}{*}{$-2446.991$} & \multirow{2}{*}{4907.981} & \multirow{2}{*}{5.166} & \multirow{2}{*}{0.076} & \multirow{2}{*}{7} \\
    &      & 2 &  1.889 & 0.609 & $-1.351$ & --- & --- & 0.700 & & & & & \\
  \hline
    &      & 1 & $-0.372$ & 0.505 &   0.019   & --- & --- & 0.217 & & & & & \\
(d) & 3-SN & 2 &   1.575  & 0.307 &   0.152   & --- & --- & 0.539 & $-2443.016$ & 4908.033 & 5.218 & 0.074 & 11 \\
    &      & 3 &   1.972  & 0.982 & $-2.219$  & --- & --- & 0.244 & & & & & \\
  \hline
\multirow{2}{*}{(e)} & \multirow{2}{*}{2-SAS} & 1 & $-0.231$ & 1.003 & --- & 0.343 & 1.237 & 0.395 & \multirow{2}{*}{$-2442.953$} & \multirow{2}{*}{4903.906} & \multirow{2}{*}{1.091} & \multirow{2}{*}{0.579} & \multirow{2}{*}{9} \\
    &       & 2 & 1.600 & 0.354 & --- & $-0.058$ & 0.872 & 0.605 & & & & & \\
  \hline
    &       & 1 & $-0.120$ & 0.575 & --- & $-0.734$ & 1.430 & 0.208 & & & & & \\
(f) & 3-SAS & 2 & $-1.192$ & 2.802 & --- &   3.365  & 4.416 & 0.409 & $-2441.530$ & 4911.060 & 8.245 & 0.016 & 14 \\
    &       & 3 &   1.592  & 0.414 & --- & $-0.036$ & 1.223 & 0.383 & & & & & \\
  \hline
\multirow{2}{*}{(g)} & \multirow{2}{*}{2-ASN} & 1 & 0.116 & 0.596 & 0.577 & --- & --- & 0.295 & \multirow{2}{*}{$-2445.935$} & \multirow{2}{*}{4905.869} & \multirow{2}{*}{3.054} & \multirow{2}{*}{0.217} & \multirow{2}{*}{7} \\
    &       & 2 & 1.199 & 0.457 & $-0.857$ & --- & --- & 0.705 & & & & & \\
  \hline
    &       & 1 & $-0.414$ & 0.771 & $-1.156$  & --- & --- & 0.059 & & & & & \\
(h) & 3-ASN & 2 &   1.701  & 0.434 &   0.403   & --- & --- & 0.646 & $-2457.621$ & 4937.243 & 34.428 & $<10^{-7}$ & 11 \\
    &       & 3 & $-0.162$ & 0.548 & $-0.031$  & --- & --- & 0.295 & & & & & \\
  \hline
\end{tabular}
\end{table*}

\begin{table*}
\caption{Parameters of the fits to the {\it Swift} data. Label corresponds to labels from Figure~\ref{fig5}. The smallest $AIC$ is marked in bold, and $p$ is the number of parameters in a model.}
\label{tbl3}
\centering
\begin{tabular}{|c c c c c c c c c c c c c c|}
\hline
  Label & Dist. & $i$ & $\mu_i$ & $\sigma_i$ & $\alpha_i$ & $\delta_i$ & $\beta_i$ & $A_i$ & $\mathcal{L}_{\rm max}$ & $AIC$ & $\Delta AIC$ & $Pr$ & $p$ \\
  \hline
\multirow{2}{*}{(a)} & \multirow{2}{*}{2-G} & 1 & $-0.026$ & 0.740 & --- & --- & --- & 0.139 & \multirow{2}{*}{$-999.848$} & \multirow{2}{*}{2009.695} & \multirow{2}{*}{14.315} & \multirow{2}{*}{0.001} & \multirow{2}{*}{5} \\
    &     & 2 &  1.638 & 0.528 & --- & --- & --- & 0.861 & & & & & \\
  \hline
    &     & 1 & $-0.435$ & 0.519 & --- & --- & --- & 0.091 & & & & & \\
(b) & 3-G & 2 &   0.875  & 0.332 & --- & --- & --- & 0.194 & $-989.654$ & \textbf{1995.308} & 0 & 1 & 8 \\
    &     & 3 &   1.785  & 0.437 & --- & --- & --- & 0.715 & & & & & \\
  \hline
\multirow{2}{*}{(c)} & \multirow{2}{*}{2-SN} & 1 & $-0.199$ & 0.622 & $-4.514$ & --- & --- & 0.059 & \multirow{2}{*}{$-991.112$} & \multirow{2}{*}{1996.348} & \multirow{2}{*}{1.040} & \multirow{2}{*}{0.632} & \multirow{2}{*}{7} \\
    &      & 2 &  2.208 & 0.915 & $-2.327$ & --- & --- & 0.941 & & & & & \\
  \hline
    &      & 1 & $-0.424$ & 0.519 & $-0.026$  & --- & --- & 0.091 & & & & & \\
(d) & 3-SN & 2 &   0.890  & 0.332 & $-0.054$  & --- & --- & 0.194 & $-989.654$ & 2001.308 & 6.000 & 0.050 & 11 \\
    &      & 3 &   1.776  & 0.437 &   0.026   & --- & --- & 0.715 & & & & & \\
  \hline
\multirow{2}{*}{(e)} & \multirow{2}{*}{2-SAS} & 1 & $-0.271$ & 0.435 & --- & $-1.044$ & 1.364 & 0.057 & \multirow{2}{*}{$-989.692$} & \multirow{2}{*}{1997.385} & \multirow{2}{*}{2.077} & \multirow{2}{*}{0.354} & \multirow{2}{*}{9} \\
    &       & 2 & 1.790 & 0.539 & --- & $-0.311$ & 0.942 & 0.943 & & & & & \\
  \hline
    &       & 1 & $-0.397$ & 0.435 & --- & $-0.386$ & 1.072 & 0.068 & & & & & \\
(f) & 3-SAS & 2 &   0.808  & 1.085 & --- &   0.801  & 1.687 & 0.250 & $-988.293$ & 2004.586 & 9.278 & 0.010 & 14 \\
    &       & 3 &   1.861  & 0.395 & --- & $-0.334$ & 0.823 & 0.682 & & & & & \\
  \hline
\multirow{2}{*}{(g)} & \multirow{2}{*}{2-ASN} & 1 & 0.126 & 0.503 & $3.035\times 10^6$ & --- & --- & 0.134 & \multirow{2}{*}{$-994.295$} & \multirow{2}{*}{2002.590} & \multirow{2}{*}{7.282} & \multirow{2}{*}{0.262} & \multirow{2}{*}{7} \\
    &       & 2 & 1.244 & 0.535 & $-1.028$ & --- & --- & 0.866 & & & & & \\
  \hline
    &       & 1 & $-0.583$ & 0.957 & $-1.091$  & --- & --- & 0.024 & & & & & \\
(h) & 3-ASN & 2 &   1.516  & 0.523 & $-0.252$  & --- & --- & 0.821 & $-1001.719$ & 2025.438 & 30.130 & $<10^{-6}$ & 11 \\
    &       & 3 & $-0.017$ & 0.887 & $-0.277$  & --- & --- & 0.155 & & & & & \\
  \hline
\end{tabular}
\end{table*}

\section{Discussion}\label{disc}

Since \citep{horvath98}, fitting a mixture of standard (i.e., non-skewed) Gaussians to the duration distribution of GRBs is a common practice. Nearly all of the catalogs examined showed that a three-Gaussian fit is statistically more significant than a two-Gaussian. This has been the basis of justifying the possibility of a third, intermediate in duration, class of GRBs, but might be ascribed simply to a higher flexibility of the fitted PDF due to a noticeably higher number of parameters. In many works, a model consisting of three Gaussians was called a trimodal, what is incorrect, as a trimodal distribution is characterized by three modes, hence three peaks recognized through local maxima \citep{schill}. This was the case only in the BATSE 3B dataset \citep{horvath98}, where 797 GRBs were examined. However, in BATSE current catalog ($\sim 2000$ GRBs) no such structure was detected \citep{horvath02,zitouni} -- it appears that the peak related to an intermediate class was smeared out when more data was gathered. Other catalogs, e.g. {\it Swift}, also exhibit a bimodal distribution, although apparently skewed. The presumed intermediate class was proposed to be linked to X-ray flares, or are related to long GRBs through some physically meaningful parameters or set of parameters \citep{veres}. Recently it was suggested \citep{zitouni} that the duration distribution corresponding to the collapsar scenario (associated to long GRBs) might not be necessary symmetric, its reason being a non-symmetric distribution of envelope masses of the progenitors. Therefore, mixtures of skewed distributions were tested herein, and it was found that a 2-SN (having the minimal $AIC$) and 2-SAS distributions are the best candidates to describe the observed $\log T_{90}$ distribution in the {\it Fermi} sample. These two models yield $\Delta_{\rm 2-SAS}<2$, which implies a substantial support for the 2-SAS model compared to a 2-SN model \cite{burnham}, corresponding to a probability of 37.7\%. Nevertheless, both of these two most plausible models are a mixture of only two skewed components. The model with the third smallest $AIC$ is a 3-G with $\Delta_{\rm 3-G}=3.119$, which gives strong support for the \mbox{3-G} model, although somewhat weaker than the preferred 2-SN and 2-SAS. The corresponding likelihood of the 3-G model is 21\%. The model with the fourth smallest $AIC$ is a 2-G, with $\Delta_{\rm 2-G}=4.459$, which means considerably less support, corresponding to a likelihood of 10.8\%. Other models yielded probabilities not higher than 3.6\%, hence are unlikely to describe the data well.

In the case of BATSE and {\it Swift}, the results are slightly different. The best model for describing their duration distribution is indeed a 3-G, however a strong support ($\Delta AIC\approx 1$) for a \mbox{2-SAS} and a 2-SN distributions indicates that models with two skewed components cannot be ruled out, although despite being of complexity comparable to the 3-G distribution (i.e., having one parameter more and one less than the 3-G), they do not introduce a third component that might be thought to come from a third class. Hence, these two-component models are of simpler interpretation, especially when the possibility that the distribution of envelope masses is non-symmetric is considered. Moreover, for the BATSE dataset, a 2-ASN and a 2-G are models with the third and fourth lowest $AIC$, with a relative probability of 21.7\% and 14.6\%, respectively. For {\it Swift}, a 2-SAS has a favorable $\Delta AIC\approx 2$, while a 2-ASN yielded a relative probability of 26.2\%, both being a considerable support. In all cases, the distributions fitted are bimodal, hence the existence of a third, intermediate in duration, GRB class is unlikely to be present in these catalogs, as well as in the {\it Fermi} sample.

It is important to note that in {\it Fermi} the sensitivity at very soft and very hard GRBs was higher than in BATSE \citep{meegan09}. Soft GRBs are intermediate in duration, and hard GRBs have short durations. Hence, an increase in intermediate GRBs relative to long ones might be expected as a consequence of improving instruments, yet the third class remains elusive \citep[e.g.][]{tarnopolski}. {\it Swift} is more sensitive in soft bands than BATSE was, hence its dataset has a low fraction of short GRBs. Therefore, the group populations inferred from {\it Fermi} observations are reasonable considering the characteristics of the instruments.

\section{Conclusions}\label{conc}

Mixtures of various statistical distributions were fitted to the observed GRB durations of {\it Fermi}, BATSE and {\it Swift}. It was found, based on the Akaike information criterion, that for {\it Fermi} the most probable among the tested models is a two-component skew-normal distribution \mbox{(2-SN)}. The second most plausible, with a relative probability of 37.7\%, is a two-component sinh-arcsinh distribution (2-SAS). A three-Gaussian has a relative probability of 21\% of being correct. It is concluded that an elusive intermediate GRB class is unlikely to be present in the {\it Fermi} duration distribution, which is better described by a two-component mixture of skewed rather than symmetric distributions, hence the third class appears to be a statistical effect, and not a physical phenomenon.

For BATSE and {\it Swift} a three-Gaussian was found to describe the distributions best, however due to the small $\Delta AIC$ the preference of a 3-G over a 2-SAS and a 2-SN, respectively, is not strong enough to rule out the latter models. Also, a considerable support is shown by a 2-ASN and a 2-G in the case of BATSE, and a 2-SAS and a 2-ASN in the case of {\it Swift}. This corroborates the possibility of a non-existence of a third, intermediate GRB class, and gives evidence that the commonly applied mixture of standard normal distributions may not be a proper model, as some skewed distributions describe the data at least as well (BATSE and {\it Swift}), or considerably better ({\it Fermi}).

\section*{Acknowledgments}
The author acknowledges support in a form of a special scholarship of Marian Smoluchowski Scientific Consortium Matter-Energy-Future from KNOW funding, grant number KNOW/48/SS/PC/2015, and wishes to thank the anonymous reviewer for useful comments that lead to significant improvement of the paper.

\end{document}